\documentclass[reprint,twocolumn, 
            amsmath,amssymb,
            tightenlines,citeautoscript,
            flushbottom, balancelastpage,
            superscriptaddress, floatfix,
            ]{revtex4-1}
\usepackage[utf8]{inputenc}
\usepackage[english]{babel}
\usepackage{hyperref}
\usepackage{dcolumn}
\usepackage{bm}
\usepackage{braket}
\usepackage{mathtools}
\usepackage{hyperref}
\usepackage[dvipsnames]{xcolor}
\usepackage{graphicx}
\usepackage[capitalise]{cleveref}
\usepackage{textcomp}
\usepackage{tabularx, cellspace}
\usepackage{bold-extra}
\usepackage{bold-extra}
\usepackage{tabularx}
\usepackage[normalem]{ulem}

\usepackage[
    protrusion=true,
    activate={true,nocompatibility},
    final,
    tracking=true,
    kerning=true,
    spacing=true,
    factor=1100]{microtype}

\hypersetup{
 pdftitle={},    
 pdfauthor={Gregory Moille},     
 pdfsubject={DKS-BS},   
 pdfcreator={Gregory Moille},   
 pdfproducer={LaTeX}, 
 pdfkeywords={Version 1.0},
 colorlinks=true,       
 linkcolor=blue,          
 citecolor=blue,        
 filecolor=blue,      
 urlcolor=blue}

\newcommand{\beginsupplement}{%
 \setcounter{table}{0}
 \renewcommand{\thetable}{S\arabic{table}}%
 \setcounter{figure}{0}
 \renewcommand{\thefigure}{S\arabic{figure}}%
 \renewcommand{\thesection}{S-\arabic{section}}
}
\usepackage{titlesec}
\titleformat*{\section}{\bfseries\Large}
\titleformat*{\subsection}{\bfseries}
\usepackage{caption}

\begin{document}
\title{Synthetic Frequency Lattices from an Integrated Dispersive Multi-Color Soliton}

\author{Gr\'egory Moille}
\email{gmoille@umd.edu}
\affiliation{Joint Quantum Institute, NIST/University of Maryland, College Park, USA}
\affiliation{Microsystems and Nanotechnology Division, National Institute of Standards and Technology, Gaithersburg, USA}
\author{Curtis Menyuk}
\affiliation{University of Maryland at Baltimore County, Baltimore, MD, USA}
\author{Yanne K. Chembo}
\affiliation{Institute for Research in Electronics and Applied Physics, University of Maryland,College Park, MD, USA}
\author{Avik Dutt}
\affiliation{Department of Mechanical Engineering and Institute for Physical Science and Technology, University of Maryland, College Park, MD 20742, USA}
\author{Kartik Srinivasan}
\affiliation{Joint Quantum Institute, NIST/University of Maryland, College Park, USA}
\affiliation{Microsystems and Nanotechnology Division, National Institute of Standards and Technology, Gaithersburg, USA}
\date{\today}
\begin{abstract}

Dissipative Kerr solitons (DKSs) in optical microresonators have been intensely studied from the perspective of both fundamental nonlinear physics and portable and low power technological applications in communications, sensing, and metrology. In parallel, synthetic dimensions offer the promise of studying physical phenomena with a dimensionality beyond that imposed by geometry, and have been implemented in optics. The interplay of DKS physics with synthetic dimensions promises to unveil numerous new physical and technological insights, yet many fundamental challenges remain.  In particular, DKSs intrinsically rely on dispersion to exist while the creation of synthetic frequency lattices typically needs a dispersion-less system. We present a change of paradigm with the creation of a synthetic frequency lattice in the eigenfrequency space of a dispersive  multi-color soliton through all-optical nonlinear coupling -- compatible with octave spanning microcombs -- harnessing the interplay between the cavity dispersion and the dispersion-less nature of the DKS. We examine theoretically and experimentally the nonlinear coupling mechanism in a 1~THz repetition rate resonator and demonstrate four-wave mixing Bragg scattering between the different wavepackets forming the multi-color soliton, with the microcomb ranging over 150~THz, yielding a complex all-optical and integrated synthetic frequency lattice.

\end{abstract}

\maketitle

Dissipative Kerr solitons (DKSs) arise from spontaneous pattern formation in cavities~\cite{LugiatoPhys.Rev.Lett.1987,GodeyPhys.Rev.A2014} balancing losses/driving forces and dispersion/nonlinear phase shift~\cite{KippenbergScience2018,DiddamsScience2020}. Therefore, the dispersion of the resonator -- either anomalous for bright pulses~\cite{LeoNaturePhoton2010,HerrNaturePhoton2014} or normal for dark pulses~\cite{XueNaturePhoton2015}, and which must not be null~\cite{ChemboPhys.Rev.A2013} -- is at the core of the DKS formation, usually through the cascaded $\chi^{(3)}$ nonlinearity, and have reached octave spanning bandwidth through dispersion~\cite{BraschScience2016,YuPhys.Rev.Applied2019, LiOptica2017,MoilleNat.Commun.2021}. Beyond generation of new frequencies, nonlinearity allows for non-trivial coupling between otherwise orthogonal azimuthal modes of the resonator~\cite{LiNaturePhoton2016,RamelowPhys.Rev.Lett.2019}. Such mode coupling has enabled exploring synthetic dimensions in photonics~\cite{OzawaPhys.Rev.A2016, YuanOptica2018}, concepts previously described and explored in atomic physics and which increase the dimensionality beyond that of the physical system's~\cite{BoadaPhys.Rev.Lett.2012,CeliPhys.Rev.Lett.2014}. Indeed, creating a link between otherwise independent resonance creates a frequency lattice, increasing synthetically the dimensionality of the system from a zero-dimensional one (\textit{i.e.} independent modes) up to a dimension dictated by the degree of coupling of the modes. Synthetic frequency lattice are of fundamental importance as they can emulate quantum-like behavior~\cite{WangLightSciAppl2020, GoldmanarXiv2022}, topology~\cite{LustigNature2019,LinNatCommun2016}, and chiral transport~\cite{DuttScience2020}.

Accessing synthetic dimensions and frequency lattices within the framework of a DKS promises uncovering new physics that is thus far impossible to realize in regular microcomb systems~\cite{TusninPhys.Rev.A2020}, such as the recent observation of Bloch oscillations in a DKS system~\cite{EnglebertArXiv}. Yet, achieving synthetic dimensions in an optical resonator typically relies on two fundamental principles that are incompatible with a broadband microcomb. First, the nonlinear coupling between neighboring modes in most resonator synthetic frequency lattices relies on electro-optic modulation~\cite{TusninPhys.Rev.A2020,EnglebertArXiv,LinNatCommun2016, QinPhys.Rev.A2018, DuttNat.Commun.2019, DingPhys.Rev.Applied2019, HuOptica2020}, and is hence limited to microwave frequencies that are incompatible with octave spanning comb repetition rate that are usually close to 1~THz~\cite{YuPhys.Rev.Applied2019,LiOptica2017,SpencerNature2018}. Therefore, all-optical frequency lattices in a resonator -- previously demonstrated in waveguides~\cite{BellOptica2017,TitchenerAPLPhotonics2020, LiPhys.Rev.A2021} -- are needed. Second, DKSs fundamentally rely on the cavity dispersion to exist, yet synthetic frequency lattices assume dispersion-less systems~\cite{DuttNatCommun2022}, limiting the size of the lattice to a narrow spectral region where dispersion can be neglected~\cite{YuanAPLPhotonics2021}.

In this paper, we demonstrate that, thanks to multi-pumping of the resonator, the interplay of the dispersion-less nature of the DKS and the dispersion of the cavity allows exploring an orthogonal dimension to the broadband microcromb mode space. We show that nonlinear coupling occurs in the DKS eigenfrequency space instead of the mode space, owing to the discrepancy between group and phase rotation velocity. Each color component of the DKS travels at the same group rotation velocity thanks to strong cross-phase modulation binding~\cite{MalomedPhys.Rev.E1993,WangOptica2017,WengPhys.Rev.X2020}, yet exhibits different phase rotation velocity because of the on-resonance multi-pumping and cavity dispersion. This allows for nonlinear mixing, ergo coupling, in this otherwise hidden dimension, similar to four-wave mixing with continuous waves. We first demonstrate theoretically and experimentally how to leverage this second dimension through nonlinear mixing via optical parametric amplification in a microcomb at 1~THz repetition rate covering more than 150~THz. We then demonstrate that thanks to the nonlinear process happening in this dimension, all optical coupling can occur and we demonstrate a synthetic frequency lattice in the eigenfrequency space of the DKS.

\textbf{Theoretical Framework}: We first aim to demonstrate that the multi-pump system can be reduced to a set of coupled Lugiato Lefever Equations (LLEs) for a discrete number of wavepacket components (the different \textit{colors} of the DKS) that oscillate beneath a single, stationary envelope. Throughout the manuscript, we only consider the fundamental transverse electric mode (TE\textsubscript{0}). Let's assume as our starting point a primary pump at an optical frequency $\omega_0$ with power $F_0$ generating a DKS, with another secondary pump $F_-$ at $\omega_-$. The effect of both driving forces can be included in a single Lugiato-Lefever equation~\cite{ChemboPhys.Rev.A2013,CoenOpt.Lett.2012} which has been shown to accurately model experiments~\cite{MoilleNat.Commun.2021,ZhangNatCommun2020}:  
\begin{align}
    \partial_t a(\theta, t) &= i \sum_m D_m A_m e^{i m \theta} - \frac{\kappa}{2}a + i\gamma|a|^2a \nonumber\\&+  F_0 e^{i\omega_0 t} +  F_- e^{i\omega_- t + i\mu_-\theta }  
\end{align}

\noindent where $A_m$ is the amplitude of the m-th azimuthal mode, $D_m$ is the angular frequency of mode m in the absence of nonlinear effects and accounts for the resonator dispersion, $\kappa$ is the attenuation coefficient and is the inverse of the photon lifetime, $\gamma$ is the nonlinear Kerr coefficient, $\theta$ is the azimuthal resonator angle, $F_0$ and $\omega_0$ are the amplitude and angular frequency of the primary pump ($\omega_0=D_0+\delta\omega_0$ with $\delta\omega_0$ the pump detuning) that we label ``$0$'', and $F_-$, $\omega_-$ and, $\mu_-$ are the amplitude, the angular frequency and the relative mode number from the main pump of the secondary pump that we label ``$-$''.

\begin{figure*}
    \centering
    \includegraphics[width = \textwidth]{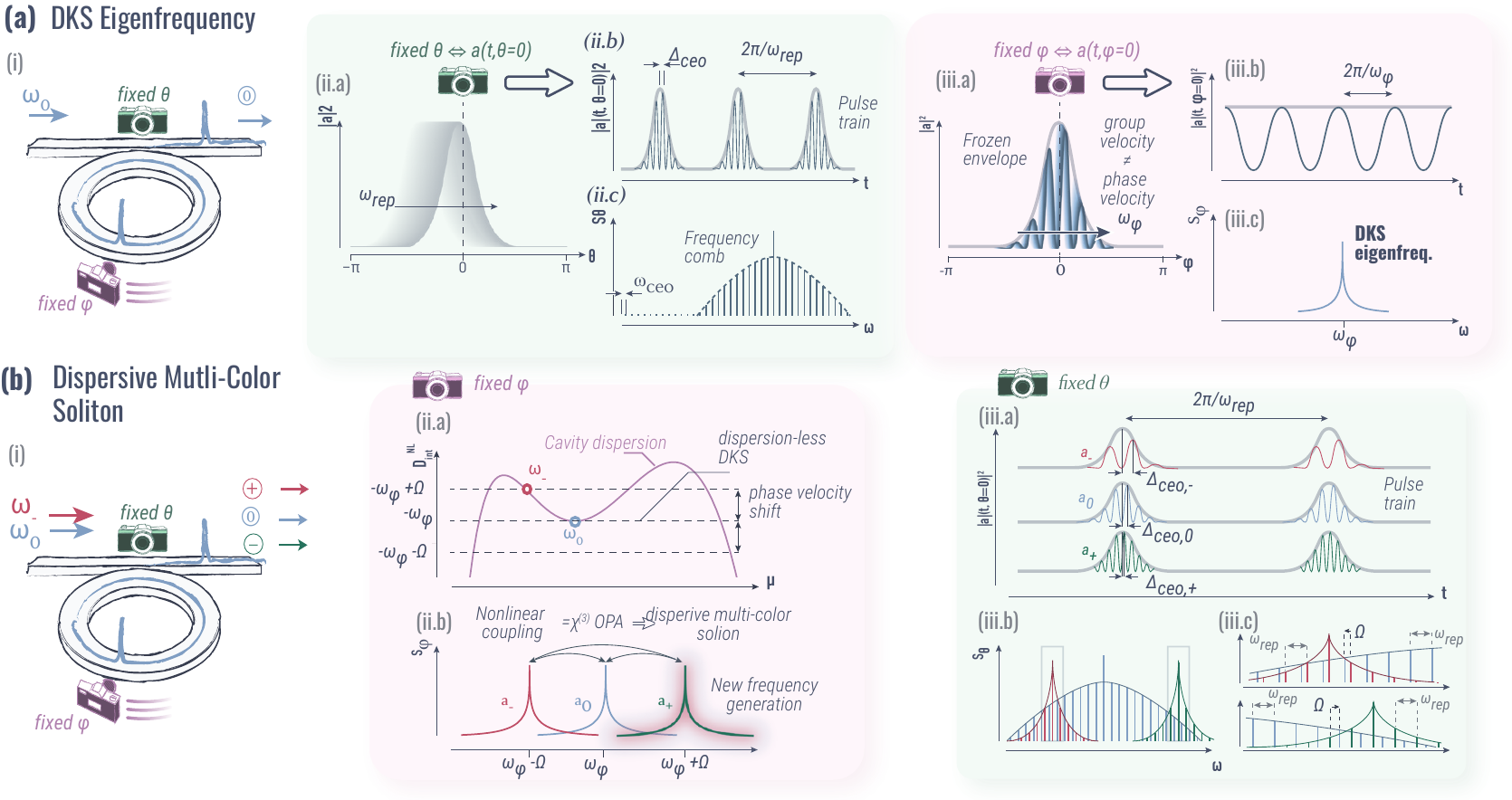}
    \caption{\label{fig:1} %
    \textbf{DKS Eigenfrequency and concept of nonlinear coupling between wavepacket components in a single resonator forming a dispersive multi-color soliton -- } %
    \textbf{(a)} Concept of soliton eigenfrequency. One could consider two points of view (i), from the waveguide (\textit{i.e} fixed $\theta$) or moving with the cavity soliton (\textit{i.e} fixed $\varphi$). In experiment, only the fixed $\theta$ viewpoint is accessible. In time, the pulse moves through this fixed point (ii.a), resulting in a pulse train in time (ii.b), which from the Fourier transform results in the frequency comb (ii.c). The phase under the envelope has a different rotation goup velocity from the rotation velocity of the soliton, and it is recorded stroboscopically, creating a fixed phase shift in the pulse train: the carrier-envelope offset. In the fixed $\varphi$ viewpoint, the envelope is effectively frozen (iii.a). Since the phase rotation velocity is different from the group velocity, the fast oscillation below the envelope travels through the fixed $\varphi$ point, resulting in a temporal profile that is a pure sine function (iii.b). It results in a single narrow line in the frequency space at $\omega_{\varphi0}$ - the eigenfrequency of the DKS. %
    \textbf{(b)} The same viewpoints can be used to examine a dual pumped DKS (i). The second external pump is located at a resonance frequency and because its mode number is different from the primary pump's mode number, the dispersion of the cavity creates a soliton component whose phase rotation velocity differs from that of the original soliton. This results in a dispersive multi-colored soliton. (ii.a). Because of strong cross phase modulation binding between these two colors in the DKS, they both move at the same group velocity. Yet the phase shift essentially shifts the secondary pump wavepacket component by the cavity dispersion at the pumped mode (ii.b). Four-wave mixing due to the third-order nonlinearity can occur, yielding an OPA-like system with an idler created at a frequency on the opposite side of the secondary pump frequency. In the coordinate system of the soliton, this idler frequency appears as a new color. Experimentally, only the fixed $\theta$ coordinate system is observable. In this framework, the difference of phase rotation velocity results in a shift of the carrier envelope offset between the three wavepacket conponents (iii.a). The frequency comb that is obtained consists of three spectrally offset components (iii.b), with the $a_{-}$ and $a_{+}$ components exhibiting an equal but opposite sign frequency shift from the central DKS frequency comb $\Omega$. The positions of the modes with resonant enhancement are defined by the phase matching and the dispersion of the resonator.
    }%
\end{figure*}  

It is usual when studying microresonator systems to make a slowly-varying envelope approximation in which it is assumed that the waveform varies slowly in a coordinate system that moves at the group velocity of the waveform.  The usual derivation of this approximation is based on the assumption that the waveform consists of a limited number of modes that in turn have a limited range of frequencies.  This assumption becomes suspect in the system that we consider here in which the mode numbers cover more than an octave.  We merely assume that the mode numbers are bounded away from zero and that the group velocities of the modes vary over a limited range due to higher-order dispersion~\cite{CherenkovPhys.Rev.A2017}.  As a result, it is possible for the Kerr nonlinearity to shift each of the mode frequencies just enough so that they lock and create a waveform that travels at a fixed rotation velocity (\textit{i.e.} fixed repetition rate $\omega_\mathrm{rep}$) -- a soliton.  Given the wide range of mode numbers, this formation of a soliton is a remarkable instance of soliton robustness~\cite{MenyukJ.Opt.Soc.Am.BJOSAB1993}.  In a system with a single pump, the rotation velocity of the soliton is largely determined by the primary pump, but in a system with a secondary pump that has a different unperturbed rotation velocity, the secondary pump pulls the rotation velocity, so that the resulting soliton rotation velocity is a combination of the two rotation velocities~\cite{TaheriEur.Phys.J.D2017}. %
This effect becomes large when the secondary pump has nearly the same magnitude as the main pump, which is the case in recent experiments~\cite{MoilleNat.Commun.2021,QureshiCommunPhys2022}. 

We define the transformation $\varphi = \theta + \omega_\mathrm{rep} t$ to effectively freeze the pulse envelope in this coordinate system. %
In the case where there is a single pump (can be generalized but chosen here for mathematical convenience), we can define $A'_m = A_m \mathrm{e}^{-i\omega_\mathrm{\varphi}t}$, in which $\omega_{\varphi0}$ is the rate of phase rotation of the soliton in the coordinate system $\varphi$ that moves with the soliton when there is a single pump and can be set equal to the eigenfrequency of the DKS [\cref{fig:1}(a)]. We then find: 

\begin{align}
    \label{eq:LLE_phi}
    \partial_t a(\varphi, t) &= i \mathcal{D} * a - \frac{\kappa}{2}a + i\gamma |a|^2a \nonumber\\&+  F_0 e^{i\delta\omega_0 t} +  F_- e^{-i\Omega t + i\mu_-\varphi} 
\end{align}
where $\mathcal{D} * a = \sum_\mu D_\mathrm{int}^{NL}(\mu) A_\mu e^{i\mu \varphi}$ with $D_\mathrm{int}^{NL} (\mu)= D_m - \left(D_0 + \omega_\mathrm{\varphi} \mu\right) = D_\mathrm{int}(\mu) - \left(\omega_\mathrm{FSR} - \omega_\mathrm{\varphi}\right)\mu$ the integrated dispersion calculated at the eigenfrequency of the DKS, $\mu = m - m_0$ the normalized mode number, $\Omega = D_\mathrm{int}^{NL}(\mu_-) + \delta\omega_-$, and $A_\mu = \int_{-\pi}^\pi a(\varphi)\mathrm{e}^{-i\mu\varphi}\mathrm{d}\varphi/2\pi$

Before turning to a discussion of what we refer in this paper as the {orthogonal dimension we are harnessing}, we will discuss in more detail the significance of the transformation that effectively `freezes' the mode-locked waveform. If we have a single soliton that in the Fourier domain corresponds to a frequency comb, then there is a striking contrast between the power spectral densities (PSDs) of $a(\theta)$ and $a(\varphi)$, referred to here respectively as $S_\theta$ and $S_\varphi$ [\cref{fig:1}(a)]. Note that we are describing the PSD of $a(\theta, t)$ at a fixed value of $\theta$ and similarly the PSD of $a(\varphi, t)$, which we would take at a point $\varphi_0$ where $a(\varphi = \varphi_0, t)$ differs significantly from zero, for instance setting $\varphi_0=0$ at the maximum of the frozen envelope [\cref{fig:1}(a.iii.a)].

In the first case, the temporal profile recreates the whole DKS pulse in a wavepacket train with periodicity $t_r = 2\pi/\omega_\mathrm{rep}$ as the pulse in the cavity moves through this fixed point [\cref{fig:1}(a.ii.a)], resulting in a pulse train in time [\cref{fig:1}(a.ii.b)]. The PSD $S_\theta$, which is the temporal Fourier transform (not to be confused with the azimuthal Fourier transform) is then a frequency comb [\cref{fig:1}(a.ii.c)].  In the second case the envelope $a(\varphi=0, t)$ is by construction constant over time. However, {the fast oscillations are not fixed under the envelope as} the phase undergoes a rotation due to the difference between the phase rotation velocity and the soliton rotation velocity that effectively `freezes' it [\cref{fig:1}(a.ii.a)]. Thus, the temporal profile of the wave at a fixed $\varphi$ is effectively a perfectly defined sine function at repetition period $2\pi/\omega_{\varphi0}$ [\cref{fig:1}(a.ii.b)].  Therefore, the PSD $S_\varphi$ is a single narrow line at the eigenfrequency of the DKS $\omega_{\varphi0}$ whose width is given by the photon lifetime [\cref{fig:1}(a.iii.c)]. 

We stress that the first case applies to every frequency comb experiment in a microesonator since every experiment extracts the DKS periodically at a fixed point in the resonator. It is also important to note that in this coordinate system the phase can slip by many factors of $2\pi$ in one round trip, but the phases are recorded stroboscopically at $\theta = 0$ (\textit{i.e.} where the DKS is extracted from the ring to the waveguide), and it is the residual slip in the phase that yields a carrier envelope offset $\omega_\mathrm{ceo}$. However, the ``frozen'' coordinate system allows for useful physical insight and theoretical understanding. 

We can use this understanding to inform our discussion of the dispersive multi-color soliton-like that forms in our experiments. 
The secondary pump at $\omega_-$, being an external driving force, must be on resonance with the cavity. We find that the secondary pump drives a wavepacket component, denoted $a_-$, that arises from FWM-BS in the mode domain~\cite{MoilleNat.Commun.2021} and that is distinct from the primary DKS. The dispersion of the cavity results in a shift of the phase rotation velocity of this wavepacket component [\cref{fig:1}(b.ii.a)]. As strong cross-phase modulation binds the wavepackets components generated by both pumps to form a single soliton that moves at a single rotation velocity~\cite{MalomedPhys.Rev.E1993, WangOptica2017}, the transformation of the azimuthal angle from $\theta$ to $\varphi$ still stands. Thus, the shift of phase rotation velocity of the secondary wave packet $a_-$ also corresponds to a narrow line shifted from the DKS eigenfrequency by the cavity dispersion and accounting for the secondary pump detuning $\Omega$, i.e., $\omega_{\varphi-} = \omega_{\varphi 0}-\Omega$ [\cref{fig:1}(b.ii.b)]. Nonlinear mixing analogous to optical parametric amplification (OPA) occurs, generating an opposite wavepacket component $a_+$ at $\omega_{\varphi+} = \omega_{\varphi0} + \Omega$.

 These frequency components at $\omega_{\varphi0} - \Omega$, $\omega_{\varphi0}$, and $\omega_{\varphi0} + \Omega$ are all distinguishable as long as their linewidths are small compared to their separation. One can then unambiguously separate $a(\varphi, t)$ into a sum of different components in the eigenfrequency dimension $\sigma$ defined by a repetition $\Omega$: 

\begin{align}
    \label{eq:three_wavepackets}
    a(\varphi, t) = \sum_{\sigma=-\infty}^{+\infty }  a_\sigma (\varphi, t)\mathrm{e}^{i(\omega_\mathrm{\varphi} +\sigma \Omega)t}
\end{align}

\vspace{2em}

Thus, this usually hidden dimension corresponds to partitioning the frequency spectrum into distinct frequency components. In the $\varphi$-domain, these components are well-defined once a soliton forms, and they are also well-defined in the $\mu$-domain. Assuming only the three wavepacket components observed in experiments~\cite{MoilleNat.Commun.2021}, re-injecting in~\cref{eq:LLE_phi}, and only keeping the components that are phase-matched, we obtain:

\begin{widetext}
        \begin{align}
            \label{eq:mutli_LLE}
            \begin{dcases}
            \partial_t a_-(\varphi, t) =  \left(-\frac{\kappa}{2} - i\Omega \right)a_- + i\mathcal{D} * a_- + i \gamma \left(|a_-|^2 + 2|a_0|^2 + 2|a_+|^2\right)a_-  + i\gamma a_0^2a_+^* + F_-\mathrm{e}^{-i\mu_-\varphi}\\
            \partial_t a_0(\varphi, t) = -\frac{\kappa}{2} a_0 + i\mathcal{D}  * a_0+i \gamma \left(2|a_-|^2 + |a_0|^2 +2|a_+|^2\right)a_0  + 2 i\gamma a_-a_+a_0^* + F_0 \\
            \partial_t a_+(\varphi,) = \left(-\frac{\kappa}{2} + i\Omega \right)a_+ + i\mathcal{D} * a_+ + i \gamma \left(2|a_-|^2 + 2|a_0|^2 +|a_+|^2\right)a_+  + i\gamma a_0^2a_-^*
            \end{dcases}
        \end{align}
    \end{widetext}

Essentially, the set of equations in \cref{eq:mutli_LLE} corresponds to four-wave mixing (FWM) between three distinct wavepacket components in the eigenfrequency dimension. Because they are all traveling at the same velocity, each wavepacket component is a color component of the soliton, with a dispersive property due to the frequency shift induced by the phase rotation velocity difference. We thus refer to this nonlinear state as a dispersive multi-color soliton. Beyond self and cross-phase modulation, nonlinear coupling between the three waves happens, analogous to what occurs in an OPA system with continuous waves. The representation in the "frozen" coordinate system $\varphi$ makes possible an intuitive understanding of the equations and further helps in understanding when more complex long-range coupling occurs, as we will later describe. However, it is important to recall that experimentally only $a(\theta  = 0, t)$ is accessible. The frequency comb is defined by three spectral envelopes traveling at the same velocity but with each shifted in phase due to the discrepancy of their phase rotation velocities in the resonator [\cref{fig:1}(b.ii.a)]. This discrepancy results in a difference of carrier envelope offsets $\Delta_\mathrm{ceo-}, \Delta_\mathrm{ceo0}, \Delta_\mathrm{ceo+}$ of the three frequency components of the wavepacket with respect to each other after one round trip (\cref{fig:1}(b.iii.a)). Given the group rotation velocity (\textit{i.e.} $\omega_\mathrm{rep}$) is the same for each colors, this leads to a constant interleaving frequency $\omega_\mathrm{ceo\pm} = \omega_\mathrm{ceo0} \mp \Omega$, which can be observed in the comb spectra [\cref{fig:1}(b.ii.b)]. Although the waveform $a(\varphi, t)$ and its spectral transform are not directly observable, the experimental demonstration of the three components traveling at the same repetition rate and their frequency shift is sufficient to verify our model. In addition, the observation of the resonance enhancement, similar to a dispersive wave, occurs at $D_\mathrm{int}^{NL} (\mu) + \sigma\Omega  =0$, and can be observed experimentally. 

\begin{figure*}
    \centering
    \includegraphics[width = \textwidth]{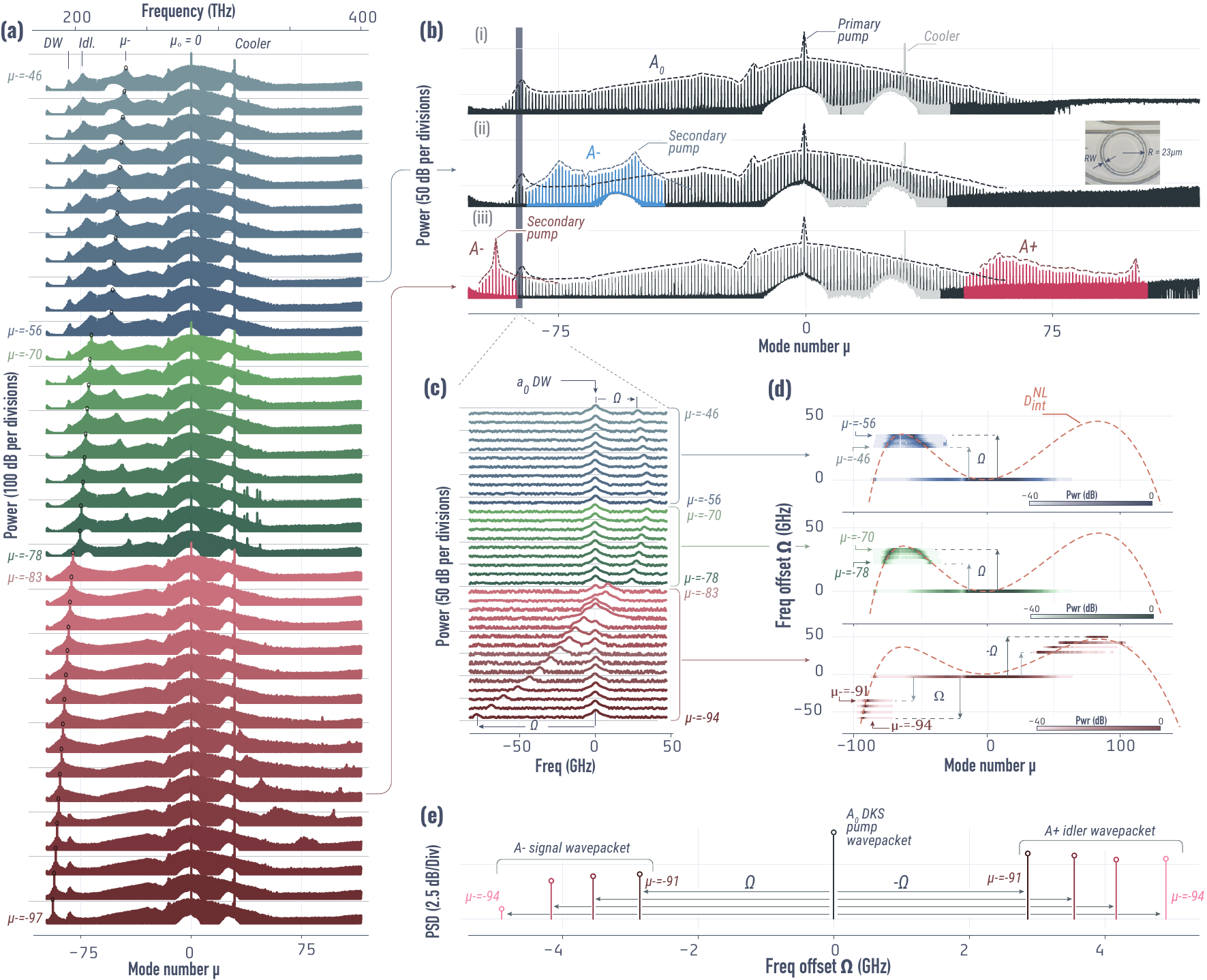}
    \caption{\label{fig:2}
    \textbf{Demonstration of nonlinear mixing forming a dispersive multi-color soliton -- }%
    \textbf{(a)} Same primary pumped DKS under different secondary pumped modes. The primary comb dispersive wave (DW), the primary pump, secondary pump, spectrally translated idler and the cooler are highlighted. The cooler is counter-propogating and cross-polarized to reduce non-linear interaction and only thermally stabilized the resonator to reach the DKS state. The colors represent different lasers used for secondary pumping, \textit{i.e.} hopping a single secondary pump mode for each spectrum of a given color, and moving discontinuously (many modes, due to gaps in laser coverage) from one color to another.  %
    \textbf{(b)} Selected frequency comb spectra highlighting the envelopes that can be resolved with an optical spectral analyzer (the counter-propagating cooler is grayed out). Panel (i) corresponds to the single pumped DKS, (ii) synthesis pump at µ =  (1307~nm), where only the $A_-$ wavepacket is observable due to coupling; (iii) secondary pump at µ = (1607~nm), where $-\Omega$ crosses the integrated dispersion, allowing observation of both $A_-$ and $A_+$ along with measurement of the spectral shift relative to $A_0$.
    \textbf{(c)} Zoom-in around the primary comb DW highlighting the variation of the spectral offset between the primary portion (with fixed DW, set at zero), and the spectrally translated portion for the pumped mode in (a). %
    \textbf{(d)} Normalized comb power (color scale) reported at their relative spectral shift. The primary pumped DKS at $\mu=0$ has its frequency shift set to be zero. The spectral shift of the translated portion follows the integrated dispersion at the secondary pump, demonstrating the same repetition rate between the wavepacket component. %
    \textbf{(e)} Power spectral density (PSD) corresponding to the bottom panel of (d), for four choices of secondary pumped mode ($\mu=-91$ to $\mu=-94$). The equally spaced tones highlight the OPA system, along with the higher power of the $A_+$ wavepacket component.}
\end{figure*}  

\textbf{Experimental Demonstration of Nonlinear Coupling Between Wavepacket Components:} We proceed to demonstrate experimentally our theory of the dispersive multi-color soliton and its nonlinear coupling in the eigenfrequency dimension. %
We use a 23~{\textmu}m radius ring resonator made of 680~nm thick silicon nitride (Si\textsubscript{3}N\textsubscript{4}) with a ring width of $RW=850$~nm embedded in silica (SiO\textsubscript{2}). The device operates with anomalous dispersion at 283~THz (1060~nm), which is the frequency of our primary pump, and enables soliton generation. We utilize a pulley-type coupler of length $L_c = 23$~{\textmu}m, which enables efficient coupling over a broad frequency band~\cite{MoilleOpt.Lett.2019}. The DKS stabilization is obtained through active cooling using a counter-propagating and cross-polarized~\cite{ZhouLightSciAppl2019} laser around 306~THz (973~nm), and whose participation in the spectral translation process can be neglected. A secondary laser is introduced and tuned on resonance for the mode from $\mu = -46$ to $\mu = -97$, while maintaining the exact same DKS state. We used three different secondary lasers (as shown in \cref{fig:2}(a)). Due to the limited tuning range of the lasers, we were not able to repeat the measurement for all modes in this range. Since the interleaving frequency $\Omega$ is larger than the resolution of the optical spectrum analyzer (20~pm, \textit{i.e.} 2.6~GHz at 200~THz and 8.17~GHz at 350~THz), we are able to segregate the envelope of each component. The primary DKS frequency comb remains unchanged from its single pumping operation and exhibits a dispersive wave (DW) component at $\mu = -86$ (about 193~THz) [\cref{fig:2}(b)]. Focusing at frequencies around the DW component, which remains fixed and independent of the secondary pumped mode, we are able to extract the interleaving frequency $\Omega$, directly related to the phase rotation velocity offset of the wavepacket components [\cref{fig:2}(c)]. %
Using this measured frequency offset, and extracting the frequency comb envelope of each comb component (\textit{i.e.} extracting $S_\theta$) from \cref{fig:2}(a), one can reproduce the two-dimensional mode number/offset frequency space of the wavepackets [\cref{fig:2}(d)]. Using the measured cavity dispersion (see supplementary material), one can infer that the common rotation frequency is $\omega_{\varphi0} \approx 981.098$~GHz. Here we report the phase rotation velocity offset as a function of the mode number for each group of modes that are continuous (\textit{i.e.} within the range of the same laser) with their corresponding colors. The plot of the frequency offset against mode number for each of the wavepacket components -- essentially highlighting the dispersive multicolor nature of the soliton -- makes it possible for us to verify the predictions of the theory. First, the repetition rate of each wavepacket component is the same. Thus, all the wavepacket components are locked together and form a single solitary wavepacket. This result can be inferred from the slopes of the wavepacket components, which are all parallel. Secondly, the resonance enhancement at the origin of the DWs, which is predicted to happen for the condition $D_\mathrm{int}^{NL} (\mu) + \sigma\Omega  =0$, occurs for each wavepacket component, reinforcing the strong binding of their group rotation velocity and highlighting the phase rotation velocity shift between them. We point out that for both 1300~nm and 1440~nm lasers (illustrated in teal and green color), we did not observe the expected OPA idler, which we believe is due to inadequate waveguide extraction of the intracavity field. In particular, the phase-matching condition of the OPA idler would occur at an even larger $\mu$ than the short wavelength DW of the original DKS, which we do not extract efficiently. However, using secondary pumping in the range $\mu_- = -91$ to $-94$, the idler ($A_{+}$) wavepacket component is observable [\cref{fig:2}(a)]. This result demonstrates the third point of the theory, which is that an idler is created with a phase shift that has the opposite sign with respect to the primary pump as does the component created by the secondary pump. In our case, we measure a frequency offset equal but of opposite sign between the DKS and $A_-(\mu)$ and $A_+(\mu)$, which also agrees with the phase matching condition for resonant enhancement [\cref{fig:2}(d), bottom panel]. Interestingly, extracting the total power of each wavepacket through summation of their envelopes highlights the OPA-like phase shift that has the opposite sign with respect to the primary pump, as does the component created by the secondary pump with an idler at higher power than the signal [\cref{fig:2}(e)].

\begin{figure*}
    \centering
    \includegraphics[width = \textwidth]{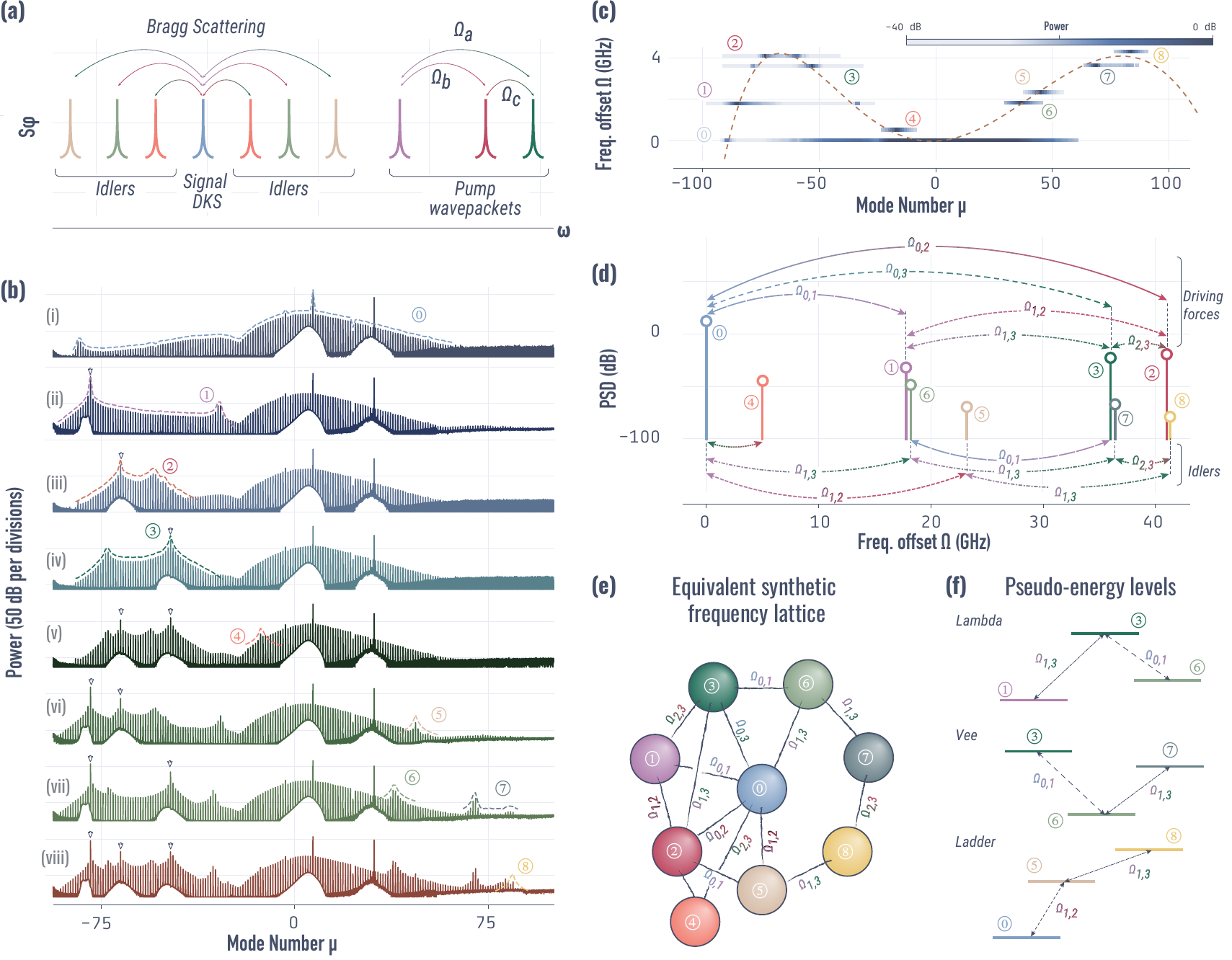}
    \caption{\label{fig:3} \textbf{Synthetic frequency lattice from DKS nonlinear mixing in the phase rotation velocity dimension -- }%
    \textbf{(a)} Four wave mixing Bragg scattering (FWM-BS) from multiple driving fields, resulting in scattering idlers from the DKS acting as the signal, with a phase shift defined by the spacing of the pumps. %
    \textbf{(b)} Optical spectra of the dissipative Kerr soliton using multiple pumps. The arrows represent the pumps. Panels (i)-(iii) show results with a single secondary pump, (iv)-(vii) show results with two secondary pumps, and (viii) shows results with three secondary pumps. The creation of new idlers is highlighted in each case by their envelopes, and their pseudo-state is numbered. %
    \textbf{(c)} Representation of the spectral envelope of the frequency comb in (a) in the two-dimensional mode number and frequency offset ($\Omega$ space (x-axis and y-axis). Each frequency offset is experimentally measured using the interleaved frequency comb. %
    \textbf{(d)} Power spectral density of each wavepacket component and their experimental frequency offset shift. The conservation of the frequency offset highlights the FWM-BS process with the idler separation driven by the secondary pump spacing. %
    \textbf{(e)} Representation of the system as a synthetic frequency lattice, where the vertices are wavepacket component and the nonlinear coupling is highlighted. %
    \textbf{(f)} Selection of pseudo-energy levels, from the definition of the eigenfrequency being the soliton photon condensate energy, that arises from the synthetic frequency lattice inside the cavity, with several configurations analogous to those found in atomic and molecular systems indicated. Here we highlight that the system presented can implement either a lambda, vee, or ladder scheme where the coupling between each set of extrema states is forbidden. It is important to note that we do not have direct access to this effective energy state as we periodically observed the intracavity field via waveguide coupling, and essentially obtain a superposition of states in the waveguide.}
\end{figure*} 

\textbf{Synthetic Frequency Lattice in the DKS Eigenfrequency Space:} %
The previously-described experiment demonstrated the possibility of harnessing soliton robustness to lock together different wavepacket components in a single wavepacket that moves at a single group rotation velocity, while each component maintains a separate phase rotation velocity, associated with the phase rotation of the secondary pump with respect to the first during one round trip. Thanks to this solitonic locking, it is possible to expand the number of driving fields allowing for four-wave mixing Bragg scattering (FWM-BS), which makes possible the creation of synthetic frequency lattices~\cite{LiOpt.Lett.2019,LiPhys.Rev.A2021, BellOptica2017,WangLightSciAppl2020}. Indeed, similar to electro-optic modulation which can be modeled as the sum of a sine component and its harmonics, FWM-BS effectively creates a moving grating at the speed of light, beating at the frequency difference between the pumps, effectively also creating a sine modulation of the refractive index. It allows for nearest neighbor modes coupling yet is not limited to low frequencies and is compatible with a large free spectral range resonator. In our system, the FWM-BS coupling happens in eigenfrequency space, and couples the different wavepacket components. Therefore, the coupling period it creates is defined by the difference of eigenfrequency between the wavepackets. For instance, if one injects three secondary pumps at three different phase rotation velocities [\cref{fig:3}(a)], three periods are imprinted, allowing for the creation of six pairs of idlers from FWM-BS of the DKS eigenfrequency. Cascaded processes can also occur, extending the range of the frequency lattice. 

We demonstrate experimentally such a synthetic frequency lattice in the phase rotation velocity space of the dispersive multi-color DKS with three secondary pumps. In order to segregate the different mixing processes, and because we do not have access to $S_\varphi$ but only $S_\theta$, we proceed to obtain every secondary pump combination. Also, because of poor out-coupling at the short wavelengths, in particular past the short DW wavelength of the DKS, as well as our inability to measure the amplitudes of the comb frequencies that fall below the frequency of the long wavelength (low-mode-number) DW, we focus only on the idlers that are generated on the positive side of $D_\mathrm{int}^\mathrm{NL}$. We start with the same single pumped DKS that we consider as our ground state  \textcircled{\footnotesize{0}} [\cref{fig:3}(b,i)]. Introducing independently a secondary pump at $\mu=-79$, $\mu=-67$, $\mu=-47$ leads to the bound wavepackets -- similar to the previous experiment -- here named \textcircled{\footnotesize{1}}, \textcircled{\footnotesize{2}}, \textcircled{\footnotesize{3}} and with a phase shift $\Omega_{0,1}$, $\Omega_{0,2}$, $\Omega_{0,3}$, respectively [\cref{fig:3}(b,ii-iv)]. When we introduce two secondary pumps together, new wavepacket components appear [\cref{fig:3}(b,v-vii)]. We later refer to these components as \textcircled{\footnotesize{4}} for the combination \{\textcircled{\footnotesize{2}},\textcircled{\footnotesize{3}}\}, \textcircled{\footnotesize{5}} for \{\textcircled{\footnotesize{1}},\textcircled{\footnotesize{2}}\}, and \textcircled{\footnotesize{6}} and \textcircled{\footnotesize{7}} for \{\textcircled{\footnotesize{1}},\textcircled{\footnotesize{3}}\}. The latter case is interesting as not one but two new wavepacket components are created, suggesting a cascaded process. Finally, injecting all the secondary pumps in the resonator simultaneously results in another new state \textcircled{\footnotesize{8}}. In the same fashion as in the previous section, and because the frequency offset between the wavepacket components is larger than the OSA resolution, we can extract both the frequency offset resulting from the phase rotation velocity offset and the spectral envelope to recreate the system two-dimensional mode number/offset frequency space~[\cref{fig:3}(c)]. Using the same DKS eigenfrequency $\omega_{\varphi0}$ one can super-impose $D_\mathrm{int}^{NL}$. This highlights how well our theoretical model predicts resonant enhancement, which is in fact closely tied to the phase shift from nonlinear mixing. In order to display the different nonlinear coupling yielding these different wavepackets, we extract the PSD, similar to the previous section, and the phase shift of each wavepacket [\cref{fig:3}(c)], further highlighting the FMW-BS process yielding a synthetic frequency lattice in the frequency offset space. As expected from the different wavepacket generation that has been segregated by the different secondary pump combinations presented in \cref{fig:3}(a), we retrieve the expected phase shift. The phase matching of the resonant enhancement modes are particularly well predicted by theory from following $D_\mathrm{int}^\mathrm{NL}$, further strengthening our confidence in the model previously described. We could recast this, as before, into a one-dimensional eigenfrequency system by extracting the PSD, to highlight the different couplings leading to the synthetic dimensions [\cref{fig:3}(d)]. Assuming the different driving field combinations yield a phase shift $\Omega_{j,k}$ with $k, j = \{1,2,3\}$, we retrieve the different FWM-BS combinations that yield new idlers. Interestingly, even though the  states \textcircled{\footnotesize{4}}, \textcircled{\footnotesize{5}} and \textcircled{\footnotesize{6}} result from scattering of the DKS state \textcircled{\footnotesize{0}}, they are further coupled to other states with a different phase shift drive. This ultimately results in a synthetic frequency lattice [\cref{fig:3}{e}] where each wavepacket is a vertex. 
Based on the theoretical prediction that soliton quantization through action-angle variables of the non-linear Schr\"odinger equation~\cite{FaddeevPhysicsReports1978, KaupJournalofMathematicalPhysics1975} results in the soliton being effectively a photon condensate at a single energy linked to $\omega_{\varphi0}$, in principle this synthetic frequency lattice is equivalent to a coupled system of different energy levels at $\hbar(\omega_{\varphi0} \pm \Omega_{j,k})$. Within this collection of pseudo-energy levels, certain couplings between states are forbidden (as they effectively do not exist). Therefore, through nonlinear coupling within a dispersive multi-color DKS, one can recreate well-known atomic configurations, such as lambda, vee and ladder state-levels [\cref{fig:3}(f)]. However, we emphasize that at this stage, our experiments remain entirely in the classical domain, and the observation of inherently quantum mechanical phenomena that have been predicted for solitons~\cite{FaddeevPhysicsReports1978} remains an outstanding challenge.


In conclusion, we have demonstrated that DKS frequency combs are not limited to their repetition rate space, and instead another orthogonal dimension is accessible through the the phase rotation velocity discrepancy between wavepackets. Similar to how OPA in a waveguide involves injecting a pump and seed field, DKS nonlinear mixing in their eigenmode space needs to be seeded by a secondary driving force to generate new wavepackets in this eigenfrequency dimension. We develop a theory for this nonlinearity resulting in a set of coupled LLEs. Essentially, the dispersion of the cavity allows for a discrepancy in the phase rotation velocity of the wavepackets -- yet bonded in group veloctiy through cross phase modulation -- allowing for nonlinear mixing and coupling in their eigenfrequency space. Through experiments using a THz repetition rate resonator, we demonstrate that all-optical nonlinear mixing in the eigenfrequency space occurs as expected by the theory presented here, where a dual-pump DKS effectively can be reduced to an OPA-like system that produces a third idler wavepacket with a phase rotation velocity offset of opposite sign from the secondary pump. Harnessing the nonlinear coupling in the phase rotation velocity space, we demonstrate the possibility to create synthetic frequency lattices in the DKS eigenfrequency space by careful selection of the secondary driving fields with the adequate phase shift caused by the cavity linear dispersion, yielding complex long range coupling. We demonstrate experimentally the creation of new idler wavepackets created through four-wave mixing Bragg-scattering, a process at the core of complex frequency lattices. The close agreement between modeling in the eigenfrequency domain and experimental observations in the laboratory frame spanning hundreds of modes over a broad frequency range firmly establishes our interpretation of the eigenfrequency space acting as a hidden dimension which can be leveraged. The presence of high-order dispersion over this broad frequency span is not a hindrance but in fact an enabler of this new synthetic dimension in DKSs.%

Our work bridges two concepts that have so far lived in different realms: synthetic dimensions through nonlinear coupling, where dispersion-less systems are sought for nearest neighbor coupling; and dissipative Kerr solitons, where dispersion is necessary to counterbalance the nonlinearity. Here, by leveraging this hidden dimension allowing for nonlinear coupling between wavepacket components, we demonstrate that dispersion can actually be harnessed to create nonlinear coupling. In addition, our demonstration relying on the four-wave mixing Bragg scattering concept allows for an all-optical synthetic dimension compatible with a $\chi^{(3)}$-only platform and high repetition rate dissipative Kerr solitons. We believe that such a first demonstration can lead the way to more complex and intricate studies of microcavity nonlinear physics. It has been shown that synthetic dimensions allow topology otherwise impossible in the linear chain of resonators. In the same fashion, we believe that our work paves the way to synthetic topology, quantum Hall effect or Bloch oscillation in the phase space, while also using the orthogonal inner high dimensionality of the frequency combs, potentially resulting in intrinsic two-dimensional complex systems. Moreover, equivalence to multi-level systems like a lambda system allows for emulation of atomic physics phenomena. Finally, the complex synthetic molecule could also be seen as a highly reconfigurable graph which could provide a versatile hardware platform for instantaneously solving graph-mapped problems for machine learning~\cite{TaitJ.LightwaveTechnol.2014,ButschekOpt.Lett.2022}.

\vspace{1ex}
\noindent \textbf{\large Data availability} \\
The data that supports the plots within this paper and other findings of this study are available from the corresponding authors upon reasonable request.\\

\vspace{1ex}
\noindent \textbf{\large Acknowledgements} \\
The ring resonators were fabricated at Ligentec Inc~\cite{NISTdisclaimer}. The authors thank Jordan Stone and Michal Chojnacky for valuable input. We also thank Kateryna Botsu for graphic design insight and Laurent Cetinsoy for fruitful discussions. The authors acknowledge funding from the DARPA APHI and NIST-on-a-chip programs. Y.K.C. acknowledges support from the Air Force Office of Scientific Research (AFOSR grant FA9550-20-1-0357).\\

\noindent \textbf{\large Author contributions}\\
G.M. led the project, designed the ring resonators, conducted the experiments and helped developed the theoretical framework. Y.K.C., A.D., and K.S. helped with the data processing and understanding. C.M. contributed in the understanding of the physical phenomenon by developing the theoretical model. G.M., C.M. and K.S. wrote the manuscript, with input from all authors. All the authors contributed and discussed the content of this manuscript.\\

\noindent \textbf{\large Competing interests}\\
The authors declare no competing interests.\\

\noindent \textbf{\large Additional Information}\\
Correspondence and requests for materials should be addressed to G.~M.

\clearpage

\beginsupplement
\section{Experimental Setup}
 
\begin{figure}[h]
    \centering
    \includegraphics[width = \columnwidth]{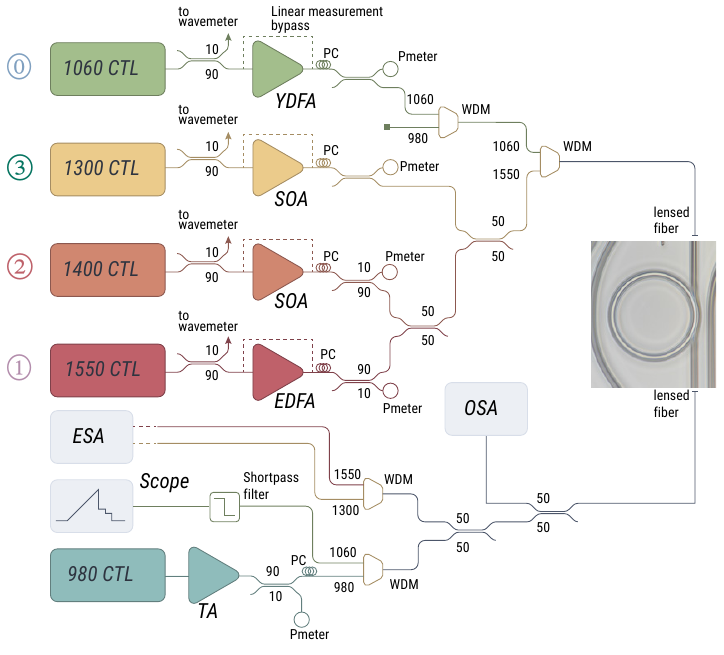}
    \caption{\label{figsup:experiment} Experimental setup used in this work. CTL: continuously tunable laser; YDFA: ytterbium doped fiber amplifier; SOA: semiconductor optical amplifier; EDFA: erbium doped fiber amplifier; TA: tapered amplifier; OSA: optical spectrum analyzer; ESA: electrical spectrum analyzer; PC: polarization controller; Pmeter: power meter; WDM: wavelength division multiplexer}
\end{figure}

 The experimental setup used in this work is depicted in \cref{figsup:experiment}, where a total of five continuously tunable laser are used, including the cooler around 980~nm. The cooler wavelength has been chosen as the resonator modes are overcoupled in this band, allowing for a large dropped power in the resonator to effectively counter the thermal bistability of the primary pump. Also, its wavelength has been chosen to be as far away as possible from the main wavelengths of interest for the nonlinear mixing processes we study. Each laser can be controlled and injected into the resonator independently, and the polarization of each laser is separately tuned to be in the transverse electric polarization once entering the on-chip waveguide. Each amplifier can be by-passed to allow for linear measurement, for instance to determine the integrated dispersion of the resonator experimentally.

\section{Integrated Dispersion Measurements}

\begin{figure}[h]
    \centering
    \includegraphics[width = \columnwidth]{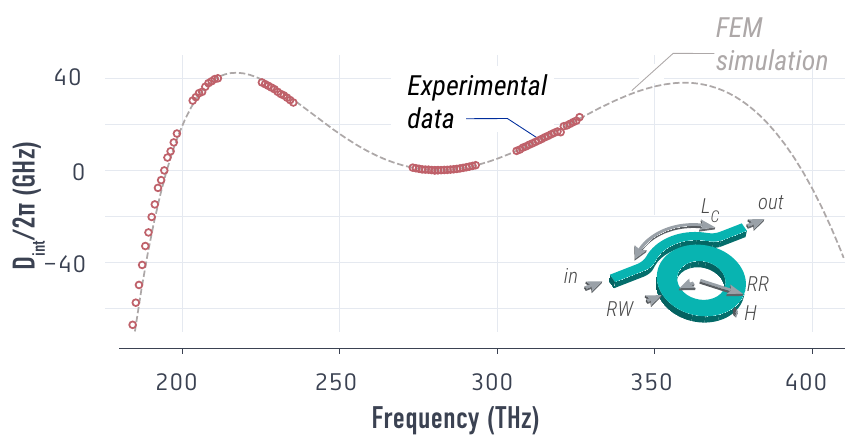}
    \caption{\label{figsup:dint} Integrated dispersion measurement (red circles) and its comparison against FEM simulations for the ring resonator studied in this work}
\end{figure}

The integrated dispersion of the resonator is crucial to our study to both understand the free spectral range along the synthetic dimension $\sigma$ and the resonant enhancement condition of each of the wavepackets. To this extent, we have proceeded to measure it carefully using all the CTLs available and referencing each resonance using a wavemeter. This allows us to retrieve the integrated dispersion of the resonator [\cref{figsup:dint}], with an uncertainty in each resonance frequency set by the wavemeter accuracy of $\approx$50~MHz. We find that the experimental integrated dispersion closely follows the expected one based on the resonator design, as computed using finite element method simulations.

\begin{thebibliography}{52}
\bibitem [{NIS()}]{NISTdisclaimer}%
      \BibitemOpen
      \href@noop {} {\enquote {\bibinfo {title} {Certain commercial products or
      names are identified to foster understanding. {{Such}} identification does
      not constitute recommendation or endorsement by the {{National Institute}} of
      {{Standards}} and {{Technology}}, nor is it intended to imply that the
      products or names identified are necessarily the best available for the
      purpose.}}\ }\BibitemShut {NoStop}%
\end{thebibliography}
\end{document}